\documentclass[
]{ceurart}

\usepackage{listings}
\usepackage{xcolor}
\usepackage{tcolorbox}

\definecolor{lightgray}{gray}{0.95}

\lstdefinestyle{promptstyle}{
    backgroundcolor=\color{lightgray},
    basicstyle=\ttfamily\small,
    frame=single,
    rulecolor=\color{black},
    breaklines=true,
    columns=fullflexible,
    captionpos=b,
    showstringspaces=false,
    keepspaces=true,
    xleftmargin=0pt,
    xrightmargin=0pt,
    aboveskip=1em,
    belowskip=1em
}

\begin{document}

\copyrightyear{2025}
\copyrightclause{Copyright for this paper by its authors.
  Use permitted under Creative Commons License Attribution 4.0
  International (CC BY 4.0).}

\conference{CLEF 2025 Working Notes, 9 -- 12 September 2025, Madrid, Spain}

\title{Deep Retrieval at CheckThat! 2025: Identifying Scientific Papers from Implicit Social Media Mentions via Hybrid Retrieval and Re-Ranking}
\title[mode=sub]{Notebook for the CheckThat! Lab at CLEF 2025}


\author[1,2,3]{Pascal J. Sager}[%
orcid=0000-0002-8084-2317,
email=pascaljosef.sager@uzh.ch,
]
\cormark[1]

\author[1]{Ashwini Kamaraj}[%
email=ashwini.kamaraj@uzh.ch,
]

\author[3]{Benjamin F. Grewe}[%
orcid=0000-0001-8560-2120,
email=bgrewe@ethz.ch,
]

\author[2,4]{Thilo Stadelmann}[%
orcid=0000-0002-3784-0420,
email=stdm@zhaw.ch,
]

\address[1]{University of Zurich, Rämistrasse 71, 8006 Zurich, Switzerland}
\address[2]{Centre for Artificial Intelligence, Zurich University of Applied Sciences, Technikumstrasse 71, 8401 Winterthur, Switzerland}
\address[3]{Institute of Neuroinformatics, ETH Zurich and University of Zurich, Winterthurerstrasse 190, 8057 Zurich, Switzerland}
\address[4]{European Centre for Living Technology, Dorsoduro 3246, 30123 Venice, Italy}

\cortext[1]{Corresponding author.}

\begin{abstract}
We present the methodology and results of the \emph{Deep Retrieval} team for subtask 4b of the CLEF CheckThat! 2025 competition, which focuses on retrieving relevant scientific literature for given social media posts. To address this task, we propose a hybrid retrieval pipeline that combines lexical precision, semantic generalization, and deep contextual re-ranking, enabling robust retrieval that bridges the informal-to-formal language gap.
Specifically, we combine BM25-based keyword matching with a FAISS vector store using a fine-tuned INF-Retriever-v1 model for dense semantic retrieval. BM25 returns the top $30$ candidates, and semantic search yields $100$ candidates, which are then merged and re-ranked via a large language model (LLM)-based cross-encoder.

Our approach achieves a mean reciprocal rank at $5$ (MRR@5) of $76.46$\% on the development set and $66.43$\% on the hidden test set, securing the \textbf{1st} position on the development leaderboard and ranking \textbf{3rd} on the test leaderboard (out of $31$ teams), with a relative performance gap of only $2$ percentage points compared to the top-ranked system.
We achieve this strong performance by running open-source models locally and without external training data, highlighting the effectiveness of a carefully designed and fine-tuned retrieval pipeline.
\end{abstract}


\maketitle

\section{Introduction}
\begin{figure}
  \centering
  \includegraphics[width=0.9\linewidth]{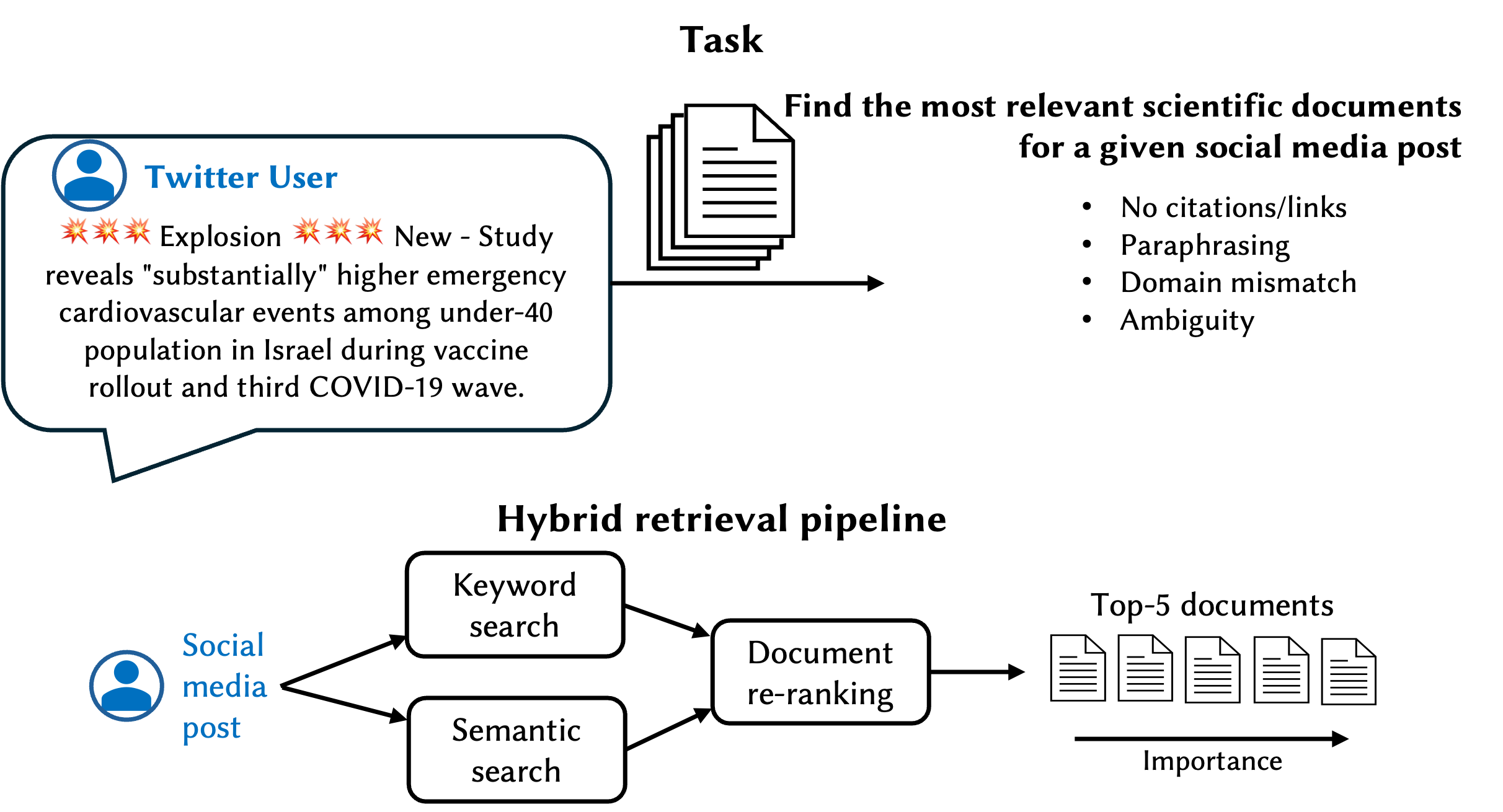}
  \caption{The \textbf{top} of the figure displays the task of finding relevant publications from a large pool of scientific documents, given a social media post. The \textbf{bottom} part of the figure provides an overview of our pipeline, consisting of three modules to predict the top five scientific documents ordered by their importance.}
  \label{fig:motivation}
\end{figure}
In the age of online misinformation, tracing social media claims back to their original scientific sources is crucial for automated fact-checking and evidence-based verification \cite{bruggemann_post-normal_2020, alam_fighting_2021}. However, this task is inherently challenging due to the linguistic and structural gap between informal, user-generated content and formal scientific literature. Social media posts often paraphrase, summarize, or loosely reference scientific findings, rarely using standardized terminology or explicit citations. These ambiguities make it difficult to reliably identify the corresponding scientific publications.

Bridging this gap requires retrieval systems that can handle domain-specific vocabulary, implicit references, and abstract semantics \cite{alam_fighting_2021}. Subtask 4b of the CLEF CheckThat! 2025 competition \cite{hauff_clef-2025_2025, clef-checkthat:2025-lncs, clef-checkthat:2025:task4} exemplifies this challenge, focusing on retrieving scientific sources for social media claims.
Figure~\ref{fig:motivation} illustrates the task and our proposed solution: A \emph{hybrid retrieval pipeline} designed specifically for cross-domain scientific source retrieval. Our method integrates:

\begin{enumerate}
    \item \textbf{Lexical retrieval} with BM25 \cite{robertson_relevance_1976, robertson_probabilistic_2009} to capture explicit term overlap (e.g., named entities, keywords);
    \item \textbf{Semantic retrieval} using a FAISS-based \cite{johnson_billion-scale_2021, douze_faiss_2024} vector store to compare dense embeddings obtained with a fine-tuned INF-Retriever-v1 \cite{infly-ai_inf-retriever-v1_nodate} model, enabling the detection of semantic overlaps;
    \item \textbf{Re-ranking} with a large language model (LLM)-based cross-encoder \cite{li_making_2023, chen_m3-embedding_2024}, which jointly encodes and scores pairs of social media posts and documents to refine relevance using deep contextual understanding.
\end{enumerate}

This architecture is designed to harness the complementary strengths of these different retrieval methods.

We evaluate our pipeline on the CheckThat! 2025 Subtask 4b dataset, achieving a Mean Reciprocal Rank at $5$ (MRR@5) of \textbf{$76.46$\%} on the development set (ranked \textbf{1st} on the leaderboard) and \textbf{$66.43$\%} on the test set (ranked \textbf{3rd} on the leaderboard out of $31$ teams), with only a $2$ percentage points lower score than the top-performing team.
\textbf{Importantly, we achieved this strong score without using any external training data, metadata, external knowledge sources, or closed-source models,} making our approach broadly applicable and easily transferable to other domains and tasks.
Overall, our main contributions are:
\begin{enumerate}
    \item A robust hybrid information retrieval (IR) architecture tailored for scientific source retrieval from informal social media content;
    \item Empirical evidence demonstrating the effectiveness of embedding fine-tuning and LLM-based re-ranking in bridging informal-to-formal domain gaps;
    \item A comprehensive experimental analysis, including ablations and a comparison to a commercial baseline.
\end{enumerate}
By publishing this well-engineered pipeline, we aim to support efforts to counter misinformation and offer a practical, open-source blueprint for cross-domain document retrieval.

\section{Related Work}
\paragraph{Fact-Checking and Scientific Source Retrieval.}
Automated fact-checking critically depends on robust document retrieval methods to identify evidence that supports or refutes a given claim \cite{thorne_fever_2018}. The evolution of this field has progressed from early strategies utilizing structured knowledge bases and curated news sources \cite{vlachos_fact_2014} to approaches that exploit unstructured, domain-specific corpora \cite{wadden_fact_2020}. A particularly challenging scenario involves retrieving scientific literature to verify claims originating from social media, due to the frequent lexical and conceptual mismatch between informal language and the academic writing style \cite{wadden_scifact-open_2022, arampatzis_overview_2023, goeuriot_overview_2024}.

\paragraph{Sparse vs. Dense Retrieval.}
Retrieval methods are commonly grouped into sparse and dense approaches.
\emph{Sparse} approaches like BM25 \cite{robertson_relevance_1976, robertson_probabilistic_2009} rely on term overlap and excel with strong lexical alignment, using probabilistic relevance frameworks with saturation parameters and document length normalization for robust ranking.
Conversely, \emph{dense} retrieval uses neural networks to encode text into vector representations, enabling semantic similarity matching through metrics such as cosine similarity \cite{karpukhin_dense_2020, izacard_leveraging_2021}. Dense models are particularly advantageous in scenarios where claims are paraphrased or loosely aligned with scientific language, as is often the case in user-generated content. 
 Although dense retrieval has historically required domain-specific fine-tuning \cite{lee_learning_2021, maillard_multi-task_2021}, recent foundation models pre-trained on diverse corpora exhibit strong generalization \cite{infly-ai_inf-retriever-v1_nodate}, increasingly blurring the distinction between general-purpose and domain-adapted retrieval.

\paragraph{Hybrid Retrieval and LLM Re-Ranking.}
Hybrid retrieval frameworks can combine sparse and dense retrieval by adding a subsequent re-ranking stage to merge their results and improve retrieval quality. Neural re-rankers \cite{nogueira_passage_2019} have demonstrated substantial improvements in ranking accuracy across multiple domains. Recently, large language models (LLMs) have been employed as cross-encoders, jointly encoding claim–document pairs to capture nuanced semantic relationships \cite{li_making_2023, chen_m3-embedding_2024}.
In this work, we adopt such a hybrid retrieval architecture by combining sparse retrieval via BM25, dense neural retrieval, and LLM-based re-ranking to leverage the different strengths of these retrieval methods.

\section{Methodology}
\begin{figure}[t]
  \centering
  \includegraphics[width=\textwidth]{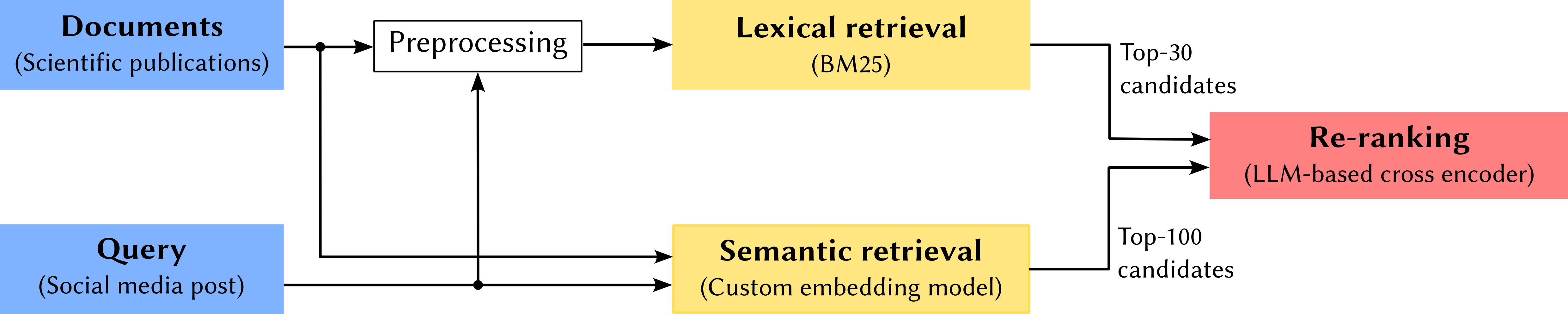}
  \caption{Overview of our retrieval pipeline. Scientific documents are indexed using two parallel retrieval mechanisms: A lexical retriever and a semantic retriever. Given a query (e.g., a social media post), the lexical retriever returns $30$ candidate papers, and the semantic retriever outputs its top $100$ candidates. These $130$ candidates are then re-ranked using a cross-encoder.}
  \label{fig:overview}
\end{figure}
Figure~\ref{fig:overview} illustrates our hybrid retrieval pipeline. In the official dataset, each document consists of a title, abstract, and limited metadata (e.g., authors, publication venue), but no full text. For indexing, we represent each document using only the title and abstract, concatenated in the format \texttt{[{Title} \textbackslash n {Abstract}]}, explicitly ignoring metadata fields.

The document corpus is processed along two parallel paths: One path applies lexical preprocessing followed by a BM25-based sparse retriever; the other encodes raw text into dense vectors for semantic retrieval. At query time, the social media post undergoes similar dual processing. Both retrieval branches return ranked candidate sets, which are merged and re-ranked using a cross-encoder. We describe each component in the following.

\subsection{Lexical Retrieval} \label{sec:lexical}
We use BM25 \cite{robertson_relevance_1976, robertson_probabilistic_2009} for sparse retrieval, and rank documents based on n-gram overlap and frequency statistics. Lexical methods are particularly effective for matching query terms to titles and commonly used scientific expressions.

\paragraph{Pre-Processing.}
In contrast to the baseline BM25 provided by the challenge organizers \cite{clef-checkthat:2025:task4}, we apply additional normalization steps to improve match quality. Our pipeline includes lowercasing, punctuation removal, and subword tokenization using byte pair encoding (BPE) \cite{sennrich_neural_2016}. We chose subword tokenization over lemmatization to maximize n-gram overlaps between informal query terms and formal document vocabulary. Hashtags are removed, while symbols such as percentages (\%) are preserved to maintain scientific meaning.
This design choice specifically addresses the domain gap between informal social media queries and formal scientific language by increasing the chance of partial term matches.
We detail additional pre-processing experiments, which were excluded from the pipeline, in Appendix \ref{sec:lexical_exp}.

\paragraph{Retrieval.} At inference time, the BM25 retriever returns the top-$30$ documents ranked by relevance. This candidate set provides strong lexical matches for downstream re-ranking and complements the semantic retriever.

\subsection{Semantic Retrieval} \label{sec:semantic}
To overcome vocabulary mismatches and paraphrasing issues, we implement dense retrieval based on transformer-derived embeddings \cite{vaswani_attention_2017}, capturing semantic similarity between queries and documents.

\paragraph{Embedding Model and Fine-tuning.}
We initialize our dense retriever with the \texttt{INF-Retriever-v1} model \cite{infly-ai_inf-retriever-v1_nodate}, a fine-tuned variant of \texttt{gte-Qwen2-7B-instruct} \cite{yang_qwen2_2024}, chosen for its strong long-text retrieval performance and open-source availability. 
We fine-tune it on the CLEF CheckThat! training set using the multiple negatives ranking (MNR) loss \cite{henderson_efficient_2017}, training it to assign higher similarity scores to (social media post, document) pairs with known associations than to random negatives.

Fine-tuning uses a maximum input length of $8,192$ tokens to avoid truncation. Queries and documents are tokenized independently, embeddings use last-token pooling, and LoRA adapters \cite{hu_lora_2022} are applied to the final eight transformer layers to reduce memory and training time \cite{tuggener_so_2024}. We optimize with AdamW \cite{loshchilov_decoupled_2019} using cosine learning rate decay and gradient accumulation. Full details of the fine-tuning setup are provided in Appendix~\ref{sec:details_embedding_ft}.

\paragraph{Vector Store.}
We pre-compute document embeddings and store them in a FAISS index \cite{johnson_billion-scale_2021, douze_faiss_2024}. The embeddings are normalized using the L2 norm, allowing cosine similarity to be computed efficiently via dot products. At inference time, the social media post is encoded into a dense vector using the same model, and the top $100$ most similar documents are retrieved. We avoid chunking abstracts, as empirical results have shown that full-document retrieval performs better.

\subsection{Re-Ranking} \label{sec:reranker}
While dense and sparse retrievers are computationally faster, the subsequent re-ranking process is computationally intensive. Unlike embedding models, which independently embed each document and query into vectors and compute similarity using a distance metric, the re-ranker processes each query-document \emph{pair} jointly to directly output a similarity score. The computational cost of these pairwise comparisons limits re-ranking to small candidate subsets, making the initial retrieval stage essential for filtering documents.

\paragraph{Ranking.}
We evaluated various re-ranking models (see Appendix~\ref{sec:rerankers}) and selected \texttt{BAAI/bge-reranker-v2-gemma} \cite{li_making_2023, chen_m3-embedding_2024}, an LLM-based cross encoder built on Gemma \cite{gemma_team_gemma_2024}, as it performed best. 
To balance cost and performance, we re-rank the top $100$ dense and top $30$ BM25 candidates, favoring the former due to stronger individual performance (see Section~\ref{sec:results}). Empirically, increasing the number of BM25 candidates beyond $30$ did not improve re-ranking performance but substantially increased computational cost, whereas increasing the dense candidates to $100$ led to substantial gains.
After removing duplicates within the $130$ candidates for each query, the re-ranker scores all remaining candidates from scratch, and the top five results are returned as output.

\section{Results} \label{sec:results}
The CLEF CheckThat! 2025 Subtask 4b evaluates systems using \emph{mean reciprocal rank at $5$ (MRR@5)}, which reflects how highly the correct source is ranked among the top five retrieved documents. Since MRR@5 is sensitive to ranking order, we prioritize optimizing the lexical and semantic retrievers for \emph{precision}.
Unlike MRR, Precision@$k$ measures the proportion of relevant documents in the top-$k$ results regardless of their order, ensuring that each retrieval stage yields high-quality candidate sets suitable for downstream re-ranking.

All experiments were conducted on the official datasets provided by the task organizers \cite{clef-checkthat:2025:task4}. The corpus includes $7,718$ documents. The development set comprises $1,400$ queries, and the test set contains $1,446$ queries.
Our complete system achieves an MRR@5 score of $76.46$\% on the development set and $66.43$\% on the test set.
Table~\ref{tab:performance_dev} summarizes development and test set results across individual and combined retrieval stages. We evaluate MRR@1 and MRR@5, along with Precision@30 and Precision@100\footnote{These metrics correspond to the best-performing configuration: BM25 returns the top 30 documents, and the semantic retriever contributes the top 100.}.
Although the absolute performance on the development set is generally higher than on the test set by approximately $10$ percentage points, the relative gains achieved through our methods, such as pre-processing and fine-tuning, are consistent across both sets.
\begin{table}
  \caption{Performance comparison on the development and test sets across retrieval and re-ranking stages. Precision is not reported for the re-ranking stage, and Elasticsearch was only evaluated on the development set. Best results in each category are in bold. ``FT'' denotes fine-tuning.}
  \label{tab:performance_dev}
  \begin{tabular}{lcc|cc|cc|cc}
    \toprule
     & \multicolumn{4}{c}{\textbf{DEVELOPMENT SET}} & \multicolumn{4}{c}{\textbf{TEST SET}}\\
     \cmidrule(lr){2-5} \cmidrule(lr){6-9}
     & \multicolumn{2}{c}{\textbf{MRR@k (\%)}} & \multicolumn{2}{c}{\textbf{Precision@k (\%)}\phantom{m}}
     & \multicolumn{2}{c}{\phantom{m}\textbf{MRR@k (\%)}} & \multicolumn{2}{c}{\textbf{Precision@k (\%)}}\\
     \cmidrule(lr){2-3} \cmidrule(lr){4-5} \cmidrule(lr){6-7} \cmidrule(lr){8-9}
      \textbf{Approach} & \textbf{k=1} & \textbf{k=5} & \textbf{k=30} & \textbf{k=100}& \textbf{k=1} & \textbf{k=5} & \textbf{k=30} & \textbf{k=100} \\
    \midrule
    \multicolumn{9}{c}{Lexical Retrieval (BM25)} \\
    \midrule
    BM25 baseline & 50.50 & 55.18 & 72.43 & 78.36 & 38.11 & 43.11 & 61.48 & 69.43\\
    \textbf{BM25 + pre-processing} & \textbf{57.50} & \textbf{62.19} & \textbf{79.86} & \textbf{85.14}  & \textbf{45.57} & \textbf{51.47} & \textbf{71.78} & \textbf{79.25}\\
    \midrule
    \multicolumn{9}{c}{Semantic Retrieval (FAISS + cosine)} \\
    \midrule
    INF-Retriever-v1 & 58.66 & 65.21 & 86.50 & 92.04  & 47.23 & 54.48 & 79.88 & 87.28\\
    \textbf{INF-Retriever-v1 + FT} & \textbf{60.86} & \textbf{67.19} & \textbf{87.86} & \textbf{93.12}  & \textbf{49.93} & \textbf{56.72} & \textbf{81.95} & \textbf{89.21}\\
    \midrule
    \multicolumn{9}{c}{Entire Pipeline} \\
    \midrule
    Elasticsearch with RRF & 63.72 & 69.35 & - & -  & - & - & - & -\\
    \textbf{Pipeline w. re-ranking} & \textbf{71.92} & \textbf{76.46} & - & -  & \textbf{60.37} & \textbf{66.43} & - & -\\
    \bottomrule
  \end{tabular}
\end{table}

\paragraph{Lexical Retrieval.}
Our BM25 retriever with additional normalization and subword tokenization yields an $8.4$-point gain in MRR@5 and a $10.3$-point gain in Precision@30 over the official baseline on the test set, similar to the improvements observed on the development set ($7.0$ points in MRR@5, $7.4$ points in Precision@30). Our preprocessing reduces noise and increases n-gram overlap, leading to better alignment between informal social media posts and formal scientific documents.

\paragraph{Semantic Retrieval.}
On the development set, employing \texttt{INF-Retriever-v1} yields an absolute improvement of $10.03$ percentage points in MRR@5 over the BM25 baseline. Fine-tuning the retriever further increases MRR@5 by $1.98$ points, reaching a final score of $67.19$\%. In terms of Precision@100, the base model achieves a $13.7$-point gain compared to the BM25 baseline, with fine-tuning contributing an additional $1.1$-point improvement. These gains are similar on the test set: \texttt{INF-Retriever-v1} improves MRR@5 by $11.4$ points over the BM25 baseline, and fine-tuning adds a further $2.2$-point gain, culminating in an MRR@5 of $56.72$\%. Precision@100 follows a similar trend, with respective gains of $17.85$ and $1.93$ percentage points. These consistent improvements across both development and test splits highlight the effectiveness and robustness of semantic retrieval, particularly when fine-tuning is applied.
We also experimented with data augmentation techniques, including HyDE-generated queries and alternative document variants. However, these did not yield further gains. A discussion on data augmentation is provided in Appendix~\ref{sec:prompt_templates}.

\paragraph{Re-Ranking.}
Our complete pipeline with \textit{bge-reranker-v2-gemma} as re-ranker achieves an MRR@5 of $76.46$\% on the development set, providing a $+9.3$ percentage points gain over our best individual retrieval method.
To isolate the effectiveness of our re-ranking approach, we compare it against a hybrid baseline using Elasticsearch (see Appendix~\ref{sec:elastic_search} for implementation details). Similar to our pipeline, this baseline uses BM25 for keyword search and the fine-tuned embedding model for semantic search, followed by reciprocal rank fusion (RRF) for re-ranking \cite{cormack_reciprocal_2009}. Although RRF provides a small boost over standalone retrieval ($+2.2$ percentage points), it underperforms the cross-encoder by $7.1$ percentage points, highlighting the added value of learning-to-rank methods.

On the test set, our pipeline achieves $66.43$\% MRR@5, with the re-ranker achieving similar improvements of $+9.7$ percentage points over the best individual retrieval method, confirming that these gains generalize across datasets.

\section{Discussion and Conclusion}
In this paper, we presented a hybrid retrieval pipeline for attributing scientific sources to social media claims. Our system combines BM25 retrieval, dense semantic search with a fine-tuned encoder, and LLM-based cross-encoder re-ranking. Our results on subtask 4b of the CLEF CheckThat! 2025 competition demonstrate the effectiveness of this architecture: We ranked \textbf{1st} on the development set and \textbf{3rd} on the test set. Key findings include:

\begin{enumerate}
\item \textbf{Hybrid retrieval is essential:}  Neither lexical nor semantic retrieval alone was sufficient. BM25 reached MRR@5 of $51.47$\%; the fine-tuned semantic retriever achieved $56.72$\%. Applying a cross-encoder to re-rank top candidates increased MRR@5 to $66.43$\% (a $23.3$ percentage point improvement over the baseline), confirming the benefit of hybrid retrieval followed by learned ranking.

\item \textbf{In-domain fine-tuning improves performance:} Fine-tuning the dense retriever improved MRR@5 by approx. $+2$ percentage points and led to a Precision@100 of $89.21$\%. While pre-trained models perform well out of the box, domain adaptation further improves alignment between informal queries and scientific abstracts.

\item \textbf{Engineering matters:}
Achieving these results required substantial engineering and experimentation efforts. We optimized hyperparameters, evaluated multiple data augmentation strategies (Appendices~\ref{sec:lexical_exp} and~\ref{sec:prompt_templates}), and evaluated alternative re-ranking models (Appendix~\ref{sec:rerankers}).
\end{enumerate}

Despite the focus on CLEF’s benchmark task, the proposed architecture is designed with broader applicability in mind. All components are modular and utilize open-source models, eliminating reliance on commercial APIs, thereby enabling deployment on local infrastructure \cite{tuggener_so_2024}. This ensures compatibility with privacy-sensitive or offline environments and facilitates customization.


\paragraph{Limitations and Future Work.}
The current pipeline does not incorporate document-level metadata, such as author names, publication venues, or timestamps, which could improve retrieval precision and disambiguation. In addition, we do not integrate external resources such as web search engines or large-scale knowledge bases. Future work could explore metadata-aware retrieval and web-based search strategies to further enhance retrieval performance.

\begin{acknowledgments}
  We thank Prof. Dr. Simon Clematide and Andrianos Michail for guiding the research, engineering, and writing process.
  We also thank the Department of Computational Linguistics at the University of Zurich and the Centre for Artificial Intelligence at the Zurich University of Applied Sciences for providing computational resources.
\end{acknowledgments}

\section*{Declaration on Generative AI}
\noindent During the creation of this work, the authors used ChatGPT\footnote{\url{https://chatgpt.com}} to refine the pre-written text. Further, the authors used Grammarly\footnote{\url{https://www.grammarly.com}} for spell checking.
 After using these tools, the authors reviewed and edited the content as needed and take full responsibility for the publication’s content. 

\bibliography{sources_cleaned}

\begin{thebibliography}{37}
\expandafter\ifx\csname natexlab\endcsname\relax\def\natexlab#1{#1}\fi
\providecommand{\url}[1]{\texttt{#1}}
\providecommand{\href}[2]{#2}
\providecommand{\path}[1]{#1}
\providecommand{\DOIprefix}{doi:}
\providecommand{\ArXivprefix}{arXiv:}
\providecommand{\URLprefix}{URL: }
\providecommand{\Pubmedprefix}{pmid:}
\providecommand{\doi}[1]{\href{http://dx.doi.org/#1}{\path{#1}}}
\providecommand{\Pubmed}[1]{\href{pmid:#1}{\path{#1}}}
\providecommand{\bibinfo}[2]{#2}
\ifx\xfnm\relax \def\xfnm[#1]{\unskip,\space#1}\fi
\bibitem[{Br\"{u}ggemann et~al.(2020)Br\"{u}ggemann, L\"{o}rcher, and Walter}]{bruggemann_post-normal_2020}
\bibinfo{author}{M.~Br\"{u}ggemann}, \bibinfo{author}{I.~L\"{o}rcher}, \bibinfo{author}{S.~Walter},
\newblock \bibinfo{title}{{Post-normal science communication: Exploring the blurring boundaries of science and journalism}},
\newblock \bibinfo{journal}{Journal of Science Communication} \bibinfo{volume}{19} (\bibinfo{year}{2020}) \bibinfo{pages}{A02}. \DOIprefix\doi{10.22323/2.19030202}.
\bibitem[{Alam et~al.(2021)Alam, Shaar, Dalvi, Sajjad, Nikolov, Mubarak et~al.}]{alam_fighting_2021}
\bibinfo{author}{F.~Alam}, \bibinfo{author}{S.~Shaar}, \bibinfo{author}{F.~Dalvi}, \bibinfo{author}{H.~Sajjad}, \bibinfo{author}{A.~Nikolov}, \bibinfo{author}{H.~Mubarak}, et~al.,
\newblock \bibinfo{title}{Fighting the {COVID}-19 {Infodemic}: {Modeling} the {Perspective} of {Journalists}, {Fact}-{Checkers}, {Social} {Media} {Platforms}, {Policy} {Makers}, and the {Society}},
\newblock in: \bibinfo{booktitle}{Findings of the {Association} for {Computational} {Linguistics}: {EMNLP} 2021}, \bibinfo{publisher}{Association for Computational Linguistics}, \bibinfo{address}{Punta Cana, Dominican Republic}, \bibinfo{year}{2021}, pp. \bibinfo{pages}{611--649}. \DOIprefix\doi{10.18653/v1/2021.findings-emnlp.56}.
\bibitem[{Alam et~al.(2025{\natexlab{a}})Alam, Stru{\ss}, Chakraborty, Dietze, Hafid, Korre, Muti, Nakov, Ruggeri, Schellhammer, Setty, Sundriyal, Todorov, and V.}]{hauff_clef-2025_2025}
\bibinfo{author}{F.~Alam}, \bibinfo{author}{J.~M. Stru{\ss}}, \bibinfo{author}{T.~Chakraborty}, \bibinfo{author}{S.~Dietze}, \bibinfo{author}{S.~Hafid}, \bibinfo{author}{K.~Korre}, \bibinfo{author}{A.~Muti}, \bibinfo{author}{P.~Nakov}, \bibinfo{author}{F.~Ruggeri}, \bibinfo{author}{S.~Schellhammer}, \bibinfo{author}{V.~Setty}, \bibinfo{author}{M.~Sundriyal}, \bibinfo{author}{K.~Todorov}, \bibinfo{author}{V.~V.},
\newblock \bibinfo{title}{{The CLEF-2025 CheckThat! Lab: Subjectivity, Fact-Checking, Claim Normalization, and Retrieval}},
\newblock in: \bibinfo{editor}{C.~Hauff}, \bibinfo{editor}{C.~Macdonald}, \bibinfo{editor}{D.~Jannach}, \bibinfo{editor}{G.~Kazai}, \bibinfo{editor}{F.~M. Nardini}, \bibinfo{editor}{F.~Pinelli}, \bibinfo{editor}{F.~Silvestri}, \bibinfo{editor}{N.~Tonellotto} (Eds.), \bibinfo{booktitle}{Advances in Information Retrieval}, \bibinfo{publisher}{Springer Nature Switzerland}, \bibinfo{address}{Cham}, \bibinfo{year}{2025}{\natexlab{a}}, pp. \bibinfo{pages}{467--478}.
\bibitem[{Alam et~al.(2025{\natexlab{b}})Alam, Struß, Chakraborty, Dietze, Hafid, Korre, Muti, Nakov, Ruggeri, Schellhammer, Setty, Sundriyal, Todorov, and Venktesh}]{clef-checkthat:2025-lncs}
\bibinfo{author}{F.~Alam}, \bibinfo{author}{J.~M. Struß}, \bibinfo{author}{T.~Chakraborty}, \bibinfo{author}{S.~Dietze}, \bibinfo{author}{S.~Hafid}, \bibinfo{author}{K.~Korre}, \bibinfo{author}{A.~Muti}, \bibinfo{author}{P.~Nakov}, \bibinfo{author}{F.~Ruggeri}, \bibinfo{author}{S.~Schellhammer}, \bibinfo{author}{V.~Setty}, \bibinfo{author}{M.~Sundriyal}, \bibinfo{author}{K.~Todorov}, \bibinfo{author}{V.~Venktesh},
\newblock \bibinfo{title}{{Overview of the CLEF-2025 CheckThat! Lab: Subjectivity, Fact-Checking, Claim Normalization, and Retrieval}},
\newblock in: \bibinfo{editor}{J.~Carrillo-de Albornoz}, \bibinfo{editor}{J.~Gonzalo}, \bibinfo{editor}{L.~Plaza}, \bibinfo{editor}{A.~García Seco~de Herrera}, \bibinfo{editor}{J.~Mothe}, \bibinfo{editor}{F.~Piroi}, \bibinfo{editor}{P.~Rosso}, \bibinfo{editor}{D.~Spina}, \bibinfo{editor}{G.~Faggioli}, \bibinfo{editor}{N.~Ferro} (Eds.), \bibinfo{booktitle}{Experimental IR Meets Multilinguality, Multimodality, and Interaction. Proceedings of the Sixteenth International Conference of the CLEF Association (CLEF 2025)}, \bibinfo{year}{2025}{\natexlab{b}}.
\bibitem[{Hafid et~al.(2025)Hafid, Kartal, Schellhammer, Boland, Dimitrov, Bringay, Todorov, and Dietze}]{clef-checkthat:2025:task4}
\bibinfo{author}{S.~Hafid}, \bibinfo{author}{Y.~S. Kartal}, \bibinfo{author}{S.~Schellhammer}, \bibinfo{author}{K.~Boland}, \bibinfo{author}{D.~Dimitrov}, \bibinfo{author}{S.~Bringay}, \bibinfo{author}{K.~Todorov}, \bibinfo{author}{S.~Dietze},
\newblock \bibinfo{title}{{Overview of the CLEF-2025 CheckThat! Lab Task 4 on Scientific Web Discourse}},
\newblock in: \bibinfo{editor}{G.~Faggioli}, \bibinfo{editor}{N.~Ferro}, \bibinfo{editor}{P.~Rosso}, \bibinfo{editor}{D.~Spina} (Eds.), \bibinfo{booktitle}{{Working Notes of CLEF 2025 - Conference and Labs of the Evaluation Forum}}, CLEF 2025, \bibinfo{address}{Madrid, Spain}, \bibinfo{year}{2025}.
\bibitem[{Robertson and Jones(1976)}]{robertson_relevance_1976}
\bibinfo{author}{S.~E. Robertson}, \bibinfo{author}{K.~S. Jones},
\newblock \bibinfo{title}{Relevance weighting of search terms},
\newblock \bibinfo{journal}{Journal of the American Society for Information Science} \bibinfo{volume}{27} (\bibinfo{year}{1976}) \bibinfo{pages}{129--146}. \DOIprefix\doi{10.1002/asi.4630270302}.
\bibitem[{Robertson and Zaragoza(2009)}]{robertson_probabilistic_2009}
\bibinfo{author}{S.~Robertson}, \bibinfo{author}{H.~Zaragoza},
\newblock \bibinfo{title}{The {Probabilistic} {Relevance} {Framework}: {BM25} and {Beyond}},
\newblock \bibinfo{journal}{Foundations and Trends\textregistered{} in Information Retrieval} \bibinfo{volume}{3} (\bibinfo{year}{2009}) \bibinfo{pages}{333--389}. \DOIprefix\doi{10.1561/1500000019}.
\bibitem[{Johnson et~al.(2021)Johnson, Douze, and Jegou}]{johnson_billion-scale_2021}
\bibinfo{author}{J.~Johnson}, \bibinfo{author}{M.~Douze}, \bibinfo{author}{H.~Jegou},
\newblock \bibinfo{title}{Billion-{Scale} {Similarity} {Search} with {GPUs}},
\newblock \bibinfo{journal}{IEEE Transactions on Big Data} \bibinfo{volume}{7} (\bibinfo{year}{2021}) \bibinfo{pages}{535--547}. \DOIprefix\doi{10.1109/TBDATA.2019.2921572}.
\bibitem[{Douze et~al.(2024)Douze, Guzhva, Deng, Johnson, Szilvasy, Mazar\'{e} et~al.}]{douze_faiss_2024}
\bibinfo{author}{M.~Douze}, \bibinfo{author}{A.~Guzhva}, \bibinfo{author}{C.~Deng}, \bibinfo{author}{J.~Johnson}, \bibinfo{author}{G.~Szilvasy}, \bibinfo{author}{P.-E. Mazar\'{e}}, et~al., \bibinfo{title}{The {Faiss} library}, \bibinfo{year}{2024}. \DOIprefix\doi{10.48550/ARXIV.2401.08281}, \bibinfo{note}{version Number: 3}.
\bibitem[{Yang et~al.(2025)Yang, Wan, Yao, Chu, Xu et~al.}]{infly-ai_inf-retriever-v1_nodate}
\bibinfo{author}{J.~Yang}, \bibinfo{author}{J.~Wan}, \bibinfo{author}{Y.~Yao}, \bibinfo{author}{W.~Chu}, \bibinfo{author}{Y.~Xu}, et~al.,
\newblock \bibinfo{title}{inf-retriever-v1}  (\bibinfo{year}{2025}). \URLprefix \url{https://huggingface.co/infly/inf-retriever-v1}. \DOIprefix\doi{10.57967/HF/4262}.
\bibitem[{Li et~al.(2023)Li, Liu, Xiao, and Shao}]{li_making_2023}
\bibinfo{author}{C.~Li}, \bibinfo{author}{Z.~Liu}, \bibinfo{author}{S.~Xiao}, \bibinfo{author}{Y.~Shao}, \bibinfo{title}{Making {Large} {Language} {Models} {A} {Better} {Foundation} {For} {Dense} {Retrieval}}, \bibinfo{year}{2023}. \DOIprefix\doi{10.48550/ARXIV.2312.15503}, \bibinfo{note}{version Number: 1}.
\bibitem[{Chen et~al.(2024)Chen, Xiao, Zhang, Luo, Lian, and Liu}]{chen_m3-embedding_2024}
\bibinfo{author}{J.~Chen}, \bibinfo{author}{S.~Xiao}, \bibinfo{author}{P.~Zhang}, \bibinfo{author}{K.~Luo}, \bibinfo{author}{D.~Lian}, \bibinfo{author}{Z.~Liu},
\newblock \bibinfo{title}{M3-{Embedding}: {Multi}-{Linguality}, {Multi}-{Functionality}, {Multi}-{Granularity} {Text} {Embeddings} {Through} {Self}-{Knowledge} {Distillation}},
\newblock in: \bibinfo{booktitle}{Findings of the {Association} for {Computational} {Linguistics} {ACL} 2024}, \bibinfo{publisher}{Association for Computational Linguistics}, \bibinfo{address}{Bangkok, Thailand and virtual meeting}, \bibinfo{year}{2024}, pp. \bibinfo{pages}{2318--2335}. \DOIprefix\doi{10.18653/v1/2024.findings-acl.137}.
\bibitem[{Thorne et~al.(2018)Thorne, Vlachos, Christodoulopoulos, and Mittal}]{thorne_fever_2018}
\bibinfo{author}{J.~Thorne}, \bibinfo{author}{A.~Vlachos}, \bibinfo{author}{C.~Christodoulopoulos}, \bibinfo{author}{A.~Mittal},
\newblock \bibinfo{title}{{FEVER}: a {Large}-scale {Dataset} for {Fact} {Extraction} and {VERification}},
\newblock in: \bibinfo{booktitle}{Proceedings of the 2018 {Conference} of the {North} {American} {Chapter} of the {Association} for {Computational} {Linguistics}: {Human} {Language} {Technologies}, {Volume} 1 ({Long} {Papers})}, \bibinfo{publisher}{Association for Computational Linguistics}, \bibinfo{address}{New Orleans, Louisiana}, \bibinfo{year}{2018}, pp. \bibinfo{pages}{809--819}. \DOIprefix\doi{10.18653/v1/N18-1074}.
\bibitem[{Vlachos and Riedel(2014)}]{vlachos_fact_2014}
\bibinfo{author}{A.~Vlachos}, \bibinfo{author}{S.~Riedel},
\newblock \bibinfo{title}{Fact {Checking}: {Task} definition and dataset construction},
\newblock in: \bibinfo{booktitle}{Proceedings of the {ACL} 2014 {Workshop} on {Language} {Technologies} and {Computational} {Social} {Science}}, \bibinfo{publisher}{Association for Computational Linguistics}, \bibinfo{address}{Baltimore, MD, USA}, \bibinfo{year}{2014}, pp. \bibinfo{pages}{18--22}. \DOIprefix\doi{10.3115/v1/W14-2508}.
\bibitem[{Wadden et~al.(2020)Wadden, Lin, Lo, Wang, Van~Zuylen, Cohan et~al.}]{wadden_fact_2020}
\bibinfo{author}{D.~Wadden}, \bibinfo{author}{S.~Lin}, \bibinfo{author}{K.~Lo}, \bibinfo{author}{L.~L. Wang}, \bibinfo{author}{M.~Van~Zuylen}, \bibinfo{author}{A.~Cohan}, et~al.,
\newblock \bibinfo{title}{Fact or {Fiction}: {Verifying} {Scientific} {Claims}},
\newblock in: \bibinfo{booktitle}{Proceedings of the 2020 {Conference} on {Empirical} {Methods} in {Natural} {Language} {Processing} ({EMNLP})}, \bibinfo{publisher}{Association for Computational Linguistics}, \bibinfo{address}{Online}, \bibinfo{year}{2020}, pp. \bibinfo{pages}{7534--7550}. \DOIprefix\doi{10.18653/v1/2020.emnlp-main.609}.
\bibitem[{Wadden et~al.(2022)Wadden, Lo, Kuehl, Cohan, Beltagy, Wang et~al.}]{wadden_scifact-open_2022}
\bibinfo{author}{D.~Wadden}, \bibinfo{author}{K.~Lo}, \bibinfo{author}{B.~Kuehl}, \bibinfo{author}{A.~Cohan}, \bibinfo{author}{I.~Beltagy}, \bibinfo{author}{L.~L. Wang}, et~al.,
\newblock \bibinfo{title}{{SciFact}-{Open}: {Towards} open-domain scientific claim verification},
\newblock in: \bibinfo{booktitle}{Findings of the {Association} for {Computational} {Linguistics}: {EMNLP} 2022}, \bibinfo{publisher}{Association for Computational Linguistics}, \bibinfo{address}{Abu Dhabi, United Arab Emirates}, \bibinfo{year}{2022}, pp. \bibinfo{pages}{4719--4734}. \DOIprefix\doi{10.18653/v1/2022.findings-emnlp.347}.
\bibitem[{Barr\'{o}n-Cede\~{n}o et~al.(2023)Barr\'{o}n-Cede\~{n}o, Alam, Galassi, Da~San~Martino, Nakov, Elsayed et~al.}]{arampatzis_overview_2023}
\bibinfo{author}{A.~Barr\'{o}n-Cede\~{n}o}, \bibinfo{author}{F.~Alam}, \bibinfo{author}{A.~Galassi}, \bibinfo{author}{G.~Da~San~Martino}, \bibinfo{author}{P.~Nakov}, \bibinfo{author}{T.~Elsayed}, et~al.,
\newblock \bibinfo{title}{Overview of the {CLEF}–2023 {CheckThat}! {Lab} on {Checkworthiness}, {Subjectivity}, {Political} {Bias}, {Factuality}, and {Authority} of {News} {Articles} and {Their} {Source}},
\newblock in: \bibinfo{booktitle}{Experimental {IR} {Meets} {Multilinguality}, {Multimodality}, and {Interaction}}, volume \bibinfo{volume}{14163}, \bibinfo{publisher}{Springer Nature Switzerland}, \bibinfo{address}{Cham}, \bibinfo{year}{2023}, pp. \bibinfo{pages}{251--275}. \DOIprefix\doi{10.1007/978-3-031-42448-9_20}, \bibinfo{note}{series Title: Lecture Notes in Computer Science}.
\bibitem[{Barr\'{o}n-Cede\~{n}o et~al.(2024)Barr\'{o}n-Cede\~{n}o, Alam, Stru\ss{}, Nakov, Chakraborty, Elsayed et~al.}]{goeuriot_overview_2024}
\bibinfo{author}{A.~Barr\'{o}n-Cede\~{n}o}, \bibinfo{author}{F.~Alam}, \bibinfo{author}{J.~M. Stru\ss{}}, \bibinfo{author}{P.~Nakov}, \bibinfo{author}{T.~Chakraborty}, \bibinfo{author}{T.~Elsayed}, et~al.,
\newblock \bibinfo{title}{Overview of the {CLEF}-2024 {CheckThat}! {Lab}: {Check}-{Worthiness}, {Subjectivity}, {Persuasion}, {Roles}, {Authorities}, and {Adversarial} {Robustness}},
\newblock in: \bibinfo{booktitle}{Experimental {IR} {Meets} {Multilinguality}, {Multimodality}, and {Interaction}}, volume \bibinfo{volume}{14959}, \bibinfo{publisher}{Springer Nature Switzerland}, \bibinfo{address}{Cham}, \bibinfo{year}{2024}, pp. \bibinfo{pages}{28--52}. \DOIprefix\doi{10.1007/978-3-031-71908-0_2}, \bibinfo{note}{series Title: Lecture Notes in Computer Science}.
\bibitem[{Karpukhin et~al.(2020)Karpukhin, Oguz, Min, Lewis, Wu, Edunov et~al.}]{karpukhin_dense_2020}
\bibinfo{author}{V.~Karpukhin}, \bibinfo{author}{B.~Oguz}, \bibinfo{author}{S.~Min}, \bibinfo{author}{P.~Lewis}, \bibinfo{author}{L.~Wu}, \bibinfo{author}{S.~Edunov}, et~al.,
\newblock \bibinfo{title}{Dense {Passage} {Retrieval} for {Open}-{Domain} {Question} {Answering}},
\newblock in: \bibinfo{booktitle}{Proceedings of the 2020 {Conference} on {Empirical} {Methods} in {Natural} {Language} {Processing} ({EMNLP})}, \bibinfo{publisher}{Association for Computational Linguistics}, \bibinfo{address}{Online}, \bibinfo{year}{2020}, pp. \bibinfo{pages}{6769--6781}. \DOIprefix\doi{10.18653/v1/2020.emnlp-main.550}.
\bibitem[{Izacard and Grave(2021)}]{izacard_leveraging_2021}
\bibinfo{author}{G.~Izacard}, \bibinfo{author}{E.~Grave},
\newblock \bibinfo{title}{Leveraging {Passage} {Retrieval} with {Generative} {Models} for {Open} {Domain} {Question} {Answering}},
\newblock in: \bibinfo{booktitle}{Proceedings of the 16th {Conference} of the {European} {Chapter} of the {Association} for {Computational} {Linguistics}: {Main} {Volume}}, \bibinfo{publisher}{Association for Computational Linguistics}, \bibinfo{address}{Online}, \bibinfo{year}{2021}, pp. \bibinfo{pages}{874--880}. \DOIprefix\doi{10.18653/v1/2021.eacl-main.74}.
\bibitem[{Lee et~al.(2021)Lee, Sung, Kang, and Chen}]{lee_learning_2021}
\bibinfo{author}{J.~Lee}, \bibinfo{author}{M.~Sung}, \bibinfo{author}{J.~Kang}, \bibinfo{author}{D.~Chen},
\newblock \bibinfo{title}{Learning {Dense} {Representations} of {Phrases} at {Scale}},
\newblock in: \bibinfo{booktitle}{Proceedings of the 59th {Annual} {Meeting} of the {Association} for {Computational} {Linguistics} and the 11th {International} {Joint} {Conference} on {Natural} {Language} {Processing} ({Volume} 1: {Long} {Papers})}, \bibinfo{publisher}{Association for Computational Linguistics}, \bibinfo{address}{Online}, \bibinfo{year}{2021}, pp. \bibinfo{pages}{6634--6647}. \DOIprefix\doi{10.18653/v1/2021.acl-long.518}.
\bibitem[{Maillard et~al.(2021)Maillard, Karpukhin, Petroni, Yih, Oguz, Stoyanov et~al.}]{maillard_multi-task_2021}
\bibinfo{author}{J.~Maillard}, \bibinfo{author}{V.~Karpukhin}, \bibinfo{author}{F.~Petroni}, \bibinfo{author}{W.-t. Yih}, \bibinfo{author}{B.~Oguz}, \bibinfo{author}{V.~Stoyanov}, et~al.,
\newblock \bibinfo{title}{Multi-{Task} {Retrieval} for {Knowledge}-{Intensive} {Tasks}},
\newblock in: \bibinfo{booktitle}{Proceedings of the 59th {Annual} {Meeting} of the {Association} for {Computational} {Linguistics} and the 11th {International} {Joint} {Conference} on {Natural} {Language} {Processing} ({Volume} 1: {Long} {Papers})}, \bibinfo{publisher}{Association for Computational Linguistics}, \bibinfo{address}{Online}, \bibinfo{year}{2021}, pp. \bibinfo{pages}{1098--1111}. \DOIprefix\doi{10.18653/v1/2021.acl-long.89}.
\bibitem[{Nogueira and Cho(2019)}]{nogueira_passage_2019}
\bibinfo{author}{R.~Nogueira}, \bibinfo{author}{K.~Cho}, \bibinfo{title}{Passage {Re}-ranking with {BERT}}, \bibinfo{year}{2019}. \DOIprefix\doi{10.48550/ARXIV.1901.04085}, \bibinfo{note}{version Number: 5}.
\bibitem[{Sennrich et~al.(2016)Sennrich, Haddow, and Birch}]{sennrich_neural_2016}
\bibinfo{author}{R.~Sennrich}, \bibinfo{author}{B.~Haddow}, \bibinfo{author}{A.~Birch},
\newblock \bibinfo{title}{Neural {Machine} {Translation} of {Rare} {Words} with {Subword} {Units}},
\newblock in: \bibinfo{booktitle}{Proceedings of the 54th {Annual} {Meeting} of the {Association} for {Computational} {Linguistics} ({Volume} 1: {Long} {Papers})}, \bibinfo{publisher}{Association for Computational Linguistics}, \bibinfo{address}{Berlin, Germany}, \bibinfo{year}{2016}, pp. \bibinfo{pages}{1715--1725}. \DOIprefix\doi{10.18653/v1/P16-1162}.
\bibitem[{Vaswani et~al.(2017)Vaswani, Shazeer, Parmar, Uszkoreit, Jones, Gomez et~al.}]{vaswani_attention_2017}
\bibinfo{author}{A.~Vaswani}, \bibinfo{author}{N.~Shazeer}, \bibinfo{author}{N.~Parmar}, \bibinfo{author}{J.~Uszkoreit}, \bibinfo{author}{L.~Jones}, \bibinfo{author}{A.~N. Gomez}, et~al.,
\newblock \bibinfo{title}{Attention is {All} you {Need}},
\newblock in: \bibinfo{booktitle}{Advances in {Neural} {Information} {Processing} {Systems}}, volume~\bibinfo{volume}{30}, \bibinfo{publisher}{Curran Associates, Inc.}, \bibinfo{year}{2017}. \URLprefix \url{https://proceedings.neurips.cc/paper_files/paper/2017/file/3f5ee243547dee91fbd053c1c4a845aa-Paper.pdf}.
\bibitem[{Yang et~al.(2024)Yang, Yang, Hui, Zheng, Yu, Zhou et~al.}]{yang_qwen2_2024}
\bibinfo{author}{A.~Yang}, \bibinfo{author}{B.~Yang}, \bibinfo{author}{B.~Hui}, \bibinfo{author}{B.~Zheng}, \bibinfo{author}{B.~Yu}, \bibinfo{author}{C.~Zhou}, et~al., \bibinfo{title}{Qwen2 {Technical} {Report}}, \bibinfo{year}{2024}. \DOIprefix\doi{10.48550/arXiv.2407.10671}, \bibinfo{note}{arXiv:2407.10671 [cs]}.
\bibitem[{Henderson et~al.(2017)Henderson, Al-Rfou, Strope, Sung, Lukacs, Guo et~al.}]{henderson_efficient_2017}
\bibinfo{author}{M.~Henderson}, \bibinfo{author}{R.~Al-Rfou}, \bibinfo{author}{B.~Strope}, \bibinfo{author}{Y.-h. Sung}, \bibinfo{author}{L.~Lukacs}, \bibinfo{author}{R.~Guo}, et~al., \bibinfo{title}{Efficient {Natural} {Language} {Response} {Suggestion} for {Smart} {Reply}}, \bibinfo{year}{2017}. \DOIprefix\doi{10.48550/arXiv.1705.00652}, \bibinfo{note}{arXiv:1705.00652 [cs]}.
\bibitem[{Hu et~al.(2022)Hu, shen, Wallis, Allen-Zhu, Li, Wang et~al.}]{hu_lora_2022}
\bibinfo{author}{E.~J. Hu}, \bibinfo{author}{y.~shen}, \bibinfo{author}{P.~Wallis}, \bibinfo{author}{Z.~Allen-Zhu}, \bibinfo{author}{Y.~Li}, \bibinfo{author}{S.~Wang}, et~al.,
\newblock \bibinfo{title}{{LoRA}: {Low}-{Rank} {Adaptation} of {Large} {Language} {Models}},
\newblock in: \bibinfo{booktitle}{International {Conference} on {Learning} {Representations}}, \bibinfo{year}{2022}. \URLprefix \url{https://openreview.net/forum?id=nZeVKeeFYf9}.
\bibitem[{Tuggener et~al.(2024)Tuggener, Sager, Taoudi-Benchekroun, Grewe, and Stadelmann}]{tuggener_so_2024}
\bibinfo{author}{L.~Tuggener}, \bibinfo{author}{P.~Sager}, \bibinfo{author}{Y.~Taoudi-Benchekroun}, \bibinfo{author}{B.~F. Grewe}, \bibinfo{author}{T.~Stadelmann},
\newblock \bibinfo{title}{So you want your private {LLM} at home? {A} survey and benchmark of methods for efficient {GPTs}},
\newblock in: \bibinfo{booktitle}{2024 11th {IEEE} {Swiss} {Conference} on {Data} {Science} ({SDS})}, \bibinfo{publisher}{IEEE}, \bibinfo{address}{Zurich, Switzerland}, \bibinfo{year}{2024}, pp. \bibinfo{pages}{205--212}. \DOIprefix\doi{10.1109/SDS60720.2024.00036}.
\bibitem[{Loshchilov and Hutter(2019)}]{loshchilov_decoupled_2019}
\bibinfo{author}{I.~Loshchilov}, \bibinfo{author}{F.~Hutter},
\newblock \bibinfo{title}{Decoupled {Weight} {Decay} {Regularization}},
\newblock in: \bibinfo{booktitle}{International {Conference} on {Learning} {Representations}}, \bibinfo{year}{2019}. \URLprefix \url{https://openreview.net/forum?id=Bkg6RiCqY7}.
\bibitem[{{Gemma Team} et~al.(2024){Gemma Team}, Riviere, Pathak, Sessa, Hardin et~al.}]{gemma_team_gemma_2024}
\bibinfo{author}{{Gemma Team}}, \bibinfo{author}{M.~Riviere}, \bibinfo{author}{S.~Pathak}, \bibinfo{author}{P.~G. Sessa}, \bibinfo{author}{C.~Hardin}, et~al., \bibinfo{title}{Gemma 2: {Improving} {Open} {Language} {Models} at a {Practical} {Size}}, \bibinfo{year}{2024}. \URLprefix \url{https://arxiv.org/abs/2408.00118}. \DOIprefix\doi{10.48550/ARXIV.2408.00118}, \bibinfo{note}{version Number: 3}.
\bibitem[{Cormack et~al.(2009)Cormack, Clarke, and Buettcher}]{cormack_reciprocal_2009}
\bibinfo{author}{G.~V. Cormack}, \bibinfo{author}{C.~L.~A. Clarke}, \bibinfo{author}{S.~Buettcher},
\newblock \bibinfo{title}{Reciprocal rank fusion outperforms condorcet and individual rank learning methods},
\newblock in: \bibinfo{booktitle}{Proceedings of the 32nd international {ACM} {SIGIR} conference on {Research} and development in information retrieval}, \bibinfo{publisher}{ACM}, \bibinfo{address}{Boston MA USA}, \bibinfo{year}{2009}, pp. \bibinfo{pages}{758--759}. \DOIprefix\doi{10.1145/1571941.1572114}.
\bibitem[{{Gemma Team} et~al.(2025){Gemma Team}, Kamath, Ferret, Pathak, Vieillard, Merhej, Perrin, Matejovicova, Ramé, Rivière, Rouillard, Mesnard, Cideron, bastien Grill, Ramos, Yvinec, Casbon, Pot, Penchev, Liu, Visin, Kenealy, Beyer, Zhai, Tsitsulin, Busa-Fekete, Feng, Sachdeva, Coleman, Gao, Mustafa, Barr, Parisotto, Tian, Eyal, Cherry, Peter, Sinopalnikov, Bhupatiraju, Agarwal, Kazemi, Malkin, Kumar, Vilar, Brusilovsky, Luo, Steiner, Friesen, Sharma, Sharma, Gilady, Goedeckemeyer, Saade, Feng, Kolesnikov, Bendebury, Abdagic, Vadi, György, Pinto, Das, Bapna, Miech, Yang, Paterson, Shenoy, Chakrabarti, Piot, Wu, Shahriari, Petrini, Chen, Lan, Choquette-Choo, Carey, Brick, Deutsch, Eisenbud, Cattle, Cheng, Paparas, Sreepathihalli, Reid, Tran, Zelle, Noland, Huizenga, Kharitonov, Liu, Amirkhanyan, Cameron, Hashemi, Klimczak-Plucińska, Singh, Mehta, Lehri, Hazimeh, Ballantyne, Szpektor, Nardini, Pouget-Abadie, Chan, Stanton, Wieting, Lai, Orbay, Fernandez, Newlan, yeong Ji, Singh, Black, Yu, Hui,
  Vodrahalli, Greff, Qiu, Valentine, Coelho, Ritter, Hoffman, Watson, Chaturvedi, Moynihan, Ma, Babar, Noy, Byrd, Roy, Momchev, Chauhan, Sachdeva, Bunyan, Botarda, Caron, Rubenstein, Culliton, Schmid, Sessa, Xu, Stanczyk, Tafti, Shivanna, Wu, Pan, Rokni, Willoughby, Vallu, Mullins, Jerome, Smoot, Girgin, Iqbal, Reddy, Sheth, Põder, Bhatnagar, Panyam, Eiger, Zhang, Liu, Yacovone, Liechty, Kalra, Evci, Misra, Roseberry, Feinberg, Kolesnikov, Han, Kwon, Chen, Chow, Zhu, Wei, Egyed, Cotruta, Giang, Kirk, Rao, Black, Babar, Lo, Moreira, Martins, Sanseviero, Gonzalez, Gleicher, Warkentin, Mirrokni, Senter, Collins, Barral, Ghahramani, Hadsell, Matias, Sculley, Petrov, Fiedel, Shazeer, Vinyals, Dean, Hassabis, Kavukcuoglu, Farabet, Buchatskaya, Alayrac, Anil, Dmitry, Lepikhin, Borgeaud, Bachem, Joulin, Andreev, Hardin, Dadashi, and Hussenot}]{gemmateam2025gemma3technicalreport}
\bibinfo{author}{{Gemma Team}}, \bibinfo{author}{A.~Kamath}, \bibinfo{author}{J.~Ferret}, \bibinfo{author}{S.~Pathak}, \bibinfo{author}{N.~Vieillard}, \bibinfo{author}{R.~Merhej}, \bibinfo{author}{S.~Perrin}, \bibinfo{author}{T.~Matejovicova}, \bibinfo{author}{A.~Ramé}, \bibinfo{author}{M.~Rivière}, \bibinfo{author}{L.~Rouillard}, \bibinfo{author}{T.~Mesnard}, \bibinfo{author}{G.~Cideron}, \bibinfo{author}{J.~bastien Grill}, \bibinfo{author}{S.~Ramos}, \bibinfo{author}{E.~Yvinec}, \bibinfo{author}{M.~Casbon}, \bibinfo{author}{E.~Pot}, \bibinfo{author}{I.~Penchev}, \bibinfo{author}{G.~Liu}, \bibinfo{author}{F.~Visin}, \bibinfo{author}{K.~Kenealy}, \bibinfo{author}{L.~Beyer}, \bibinfo{author}{X.~Zhai}, \bibinfo{author}{A.~Tsitsulin}, \bibinfo{author}{R.~Busa-Fekete}, \bibinfo{author}{A.~Feng}, \bibinfo{author}{N.~Sachdeva}, \bibinfo{author}{B.~Coleman}, \bibinfo{author}{Y.~Gao}, \bibinfo{author}{B.~Mustafa}, \bibinfo{author}{I.~Barr}, \bibinfo{author}{E.~Parisotto}, \bibinfo{author}{D.~Tian},
  \bibinfo{author}{M.~Eyal}, \bibinfo{author}{C.~Cherry}, \bibinfo{author}{J.-T. Peter}, \bibinfo{author}{D.~Sinopalnikov}, \bibinfo{author}{S.~Bhupatiraju}, \bibinfo{author}{R.~Agarwal}, \bibinfo{author}{M.~Kazemi}, \bibinfo{author}{D.~Malkin}, \bibinfo{author}{R.~Kumar}, \bibinfo{author}{D.~Vilar}, \bibinfo{author}{I.~Brusilovsky}, \bibinfo{author}{J.~Luo}, \bibinfo{author}{A.~Steiner}, \bibinfo{author}{A.~Friesen}, \bibinfo{author}{A.~Sharma}, \bibinfo{author}{A.~Sharma}, \bibinfo{author}{A.~M. Gilady}, \bibinfo{author}{A.~Goedeckemeyer}, \bibinfo{author}{A.~Saade}, \bibinfo{author}{A.~Feng}, \bibinfo{author}{A.~Kolesnikov}, \bibinfo{author}{A.~Bendebury}, \bibinfo{author}{A.~Abdagic}, \bibinfo{author}{A.~Vadi}, \bibinfo{author}{A.~György}, \bibinfo{author}{A.~S. Pinto}, \bibinfo{author}{A.~Das}, \bibinfo{author}{A.~Bapna}, \bibinfo{author}{A.~Miech}, \bibinfo{author}{A.~Yang}, \bibinfo{author}{A.~Paterson}, \bibinfo{author}{A.~Shenoy}, \bibinfo{author}{A.~Chakrabarti}, \bibinfo{author}{B.~Piot},
  \bibinfo{author}{B.~Wu}, \bibinfo{author}{B.~Shahriari}, \bibinfo{author}{B.~Petrini}, \bibinfo{author}{C.~Chen}, \bibinfo{author}{C.~L. Lan}, \bibinfo{author}{C.~A. Choquette-Choo}, \bibinfo{author}{C.~Carey}, \bibinfo{author}{C.~Brick}, \bibinfo{author}{D.~Deutsch}, \bibinfo{author}{D.~Eisenbud}, \bibinfo{author}{D.~Cattle}, \bibinfo{author}{D.~Cheng}, \bibinfo{author}{D.~Paparas}, \bibinfo{author}{D.~S. Sreepathihalli}, \bibinfo{author}{D.~Reid}, \bibinfo{author}{D.~Tran}, \bibinfo{author}{D.~Zelle}, \bibinfo{author}{E.~Noland}, \bibinfo{author}{E.~Huizenga}, \bibinfo{author}{E.~Kharitonov}, \bibinfo{author}{F.~Liu}, \bibinfo{author}{G.~Amirkhanyan}, \bibinfo{author}{G.~Cameron}, \bibinfo{author}{H.~Hashemi}, \bibinfo{author}{H.~Klimczak-Plucińska}, \bibinfo{author}{H.~Singh}, \bibinfo{author}{H.~Mehta}, \bibinfo{author}{H.~T. Lehri}, \bibinfo{author}{H.~Hazimeh}, \bibinfo{author}{I.~Ballantyne}, \bibinfo{author}{I.~Szpektor}, \bibinfo{author}{I.~Nardini}, \bibinfo{author}{J.~Pouget-Abadie},
  \bibinfo{author}{J.~Chan}, \bibinfo{author}{J.~Stanton}, \bibinfo{author}{J.~Wieting}, \bibinfo{author}{J.~Lai}, \bibinfo{author}{J.~Orbay}, \bibinfo{author}{J.~Fernandez}, \bibinfo{author}{J.~Newlan}, \bibinfo{author}{J.~yeong Ji}, \bibinfo{author}{J.~Singh}, \bibinfo{author}{K.~Black}, \bibinfo{author}{K.~Yu}, \bibinfo{author}{K.~Hui}, \bibinfo{author}{K.~Vodrahalli}, \bibinfo{author}{K.~Greff}, \bibinfo{author}{L.~Qiu}, \bibinfo{author}{M.~Valentine}, \bibinfo{author}{M.~Coelho}, \bibinfo{author}{M.~Ritter}, \bibinfo{author}{M.~Hoffman}, \bibinfo{author}{M.~Watson}, \bibinfo{author}{M.~Chaturvedi}, \bibinfo{author}{M.~Moynihan}, \bibinfo{author}{M.~Ma}, \bibinfo{author}{N.~Babar}, \bibinfo{author}{N.~Noy}, \bibinfo{author}{N.~Byrd}, \bibinfo{author}{N.~Roy}, \bibinfo{author}{N.~Momchev}, \bibinfo{author}{N.~Chauhan}, \bibinfo{author}{N.~Sachdeva}, \bibinfo{author}{O.~Bunyan}, \bibinfo{author}{P.~Botarda}, \bibinfo{author}{P.~Caron}, \bibinfo{author}{P.~K. Rubenstein}, \bibinfo{author}{P.~Culliton},
  \bibinfo{author}{P.~Schmid}, \bibinfo{author}{P.~G. Sessa}, \bibinfo{author}{P.~Xu}, \bibinfo{author}{P.~Stanczyk}, \bibinfo{author}{P.~Tafti}, \bibinfo{author}{R.~Shivanna}, \bibinfo{author}{R.~Wu}, \bibinfo{author}{R.~Pan}, \bibinfo{author}{R.~Rokni}, \bibinfo{author}{R.~Willoughby}, \bibinfo{author}{R.~Vallu}, \bibinfo{author}{R.~Mullins}, \bibinfo{author}{S.~Jerome}, \bibinfo{author}{S.~Smoot}, \bibinfo{author}{S.~Girgin}, \bibinfo{author}{S.~Iqbal}, \bibinfo{author}{S.~Reddy}, \bibinfo{author}{S.~Sheth}, \bibinfo{author}{S.~Põder}, \bibinfo{author}{S.~Bhatnagar}, \bibinfo{author}{S.~R. Panyam}, \bibinfo{author}{S.~Eiger}, \bibinfo{author}{S.~Zhang}, \bibinfo{author}{T.~Liu}, \bibinfo{author}{T.~Yacovone}, \bibinfo{author}{T.~Liechty}, \bibinfo{author}{U.~Kalra}, \bibinfo{author}{U.~Evci}, \bibinfo{author}{V.~Misra}, \bibinfo{author}{V.~Roseberry}, \bibinfo{author}{V.~Feinberg}, \bibinfo{author}{V.~Kolesnikov}, \bibinfo{author}{W.~Han}, \bibinfo{author}{W.~Kwon}, \bibinfo{author}{X.~Chen},
  \bibinfo{author}{Y.~Chow}, \bibinfo{author}{Y.~Zhu}, \bibinfo{author}{Z.~Wei}, \bibinfo{author}{Z.~Egyed}, \bibinfo{author}{V.~Cotruta}, \bibinfo{author}{M.~Giang}, \bibinfo{author}{P.~Kirk}, \bibinfo{author}{A.~Rao}, \bibinfo{author}{K.~Black}, \bibinfo{author}{N.~Babar}, \bibinfo{author}{J.~Lo}, \bibinfo{author}{E.~Moreira}, \bibinfo{author}{L.~G. Martins}, \bibinfo{author}{O.~Sanseviero}, \bibinfo{author}{L.~Gonzalez}, \bibinfo{author}{Z.~Gleicher}, \bibinfo{author}{T.~Warkentin}, \bibinfo{author}{V.~Mirrokni}, \bibinfo{author}{E.~Senter}, \bibinfo{author}{E.~Collins}, \bibinfo{author}{J.~Barral}, \bibinfo{author}{Z.~Ghahramani}, \bibinfo{author}{R.~Hadsell}, \bibinfo{author}{Y.~Matias}, \bibinfo{author}{D.~Sculley}, \bibinfo{author}{S.~Petrov}, \bibinfo{author}{N.~Fiedel}, \bibinfo{author}{N.~Shazeer}, \bibinfo{author}{O.~Vinyals}, \bibinfo{author}{J.~Dean}, \bibinfo{author}{D.~Hassabis}, \bibinfo{author}{K.~Kavukcuoglu}, \bibinfo{author}{C.~Farabet}, \bibinfo{author}{E.~Buchatskaya},
  \bibinfo{author}{J.-B. Alayrac}, \bibinfo{author}{R.~Anil}, \bibinfo{author}{Dmitry}, \bibinfo{author}{Lepikhin}, \bibinfo{author}{S.~Borgeaud}, \bibinfo{author}{O.~Bachem}, \bibinfo{author}{A.~Joulin}, \bibinfo{author}{A.~Andreev}, \bibinfo{author}{C.~Hardin}, \bibinfo{author}{R.~Dadashi}, \bibinfo{author}{L.~Hussenot}, \bibinfo{title}{Gemma 3 technical report}, \bibinfo{year}{2025}. \DOIprefix\doi{10.48550/arXiv.2503.19786}.
\bibitem[{Shakir et~al.(2024)Shakir, Koenig, Lipp, and Lee}]{rerank2024mxbai}
\bibinfo{author}{A.~Shakir}, \bibinfo{author}{D.~Koenig}, \bibinfo{author}{J.~Lipp}, \bibinfo{author}{S.~Lee}, \bibinfo{title}{{Boost Your Search With The Crispy Mixedbread Rerank Models}}, \bibinfo{year}{2024}. \URLprefix \url{https://www.mixedbread.ai/blog/mxbai-rerank-v1}.
\bibitem[{Xiao et~al.(2023)Xiao, Liu, Zhang, and Muennighoff}]{bge_embedding}
\bibinfo{author}{S.~Xiao}, \bibinfo{author}{Z.~Liu}, \bibinfo{author}{P.~Zhang}, \bibinfo{author}{N.~Muennighoff}, \bibinfo{title}{C-pack: Packaged resources to advance general chinese embedding}, \bibinfo{year}{2023}. \href{http://arxiv.org/abs/2309.07597}{{\tt arXiv:2309.07597}}.
\bibitem[{Gao et~al.(2023)Gao, Ma, Lin, and Callan}]{gao_precise_2023}
\bibinfo{author}{L.~Gao}, \bibinfo{author}{X.~Ma}, \bibinfo{author}{J.~Lin}, \bibinfo{author}{J.~Callan},
\newblock \bibinfo{title}{Precise {Zero}-{Shot} {Dense} {Retrieval} without {Relevance} {Labels}},
\newblock in: \bibinfo{booktitle}{Proceedings of the 61st {Annual} {Meeting} of the {Association} for {Computational} {Linguistics} ({Volume} 1: {Long} {Papers})}, \bibinfo{publisher}{Association for Computational Linguistics}, \bibinfo{address}{Toronto, Canada}, \bibinfo{year}{2023}, pp. \bibinfo{pages}{1762--1777}. \DOIprefix\doi{10.18653/v1/2023.acl-long.99}.
\bibitem[{Grattafiori et~al.(2024)Grattafiori, Dubey, Jauhri, Pandey, Kadian, Al-Dahle et~al.}]{grattafiori_llama_2024}
\bibinfo{author}{A.~Grattafiori}, \bibinfo{author}{A.~Dubey}, \bibinfo{author}{A.~Jauhri}, \bibinfo{author}{A.~Pandey}, \bibinfo{author}{A.~Kadian}, \bibinfo{author}{A.~Al-Dahle}, et~al., \bibinfo{title}{The {Llama} 3 {Herd} of {Models}}, \bibinfo{year}{2024}. \DOIprefix\doi{10.48550/ARXIV.2407.21783}, \bibinfo{note}{version Number: 3}.

\end{thebibliography}

\appendix
\section{Additional Experiments on Lexical Retrieval} \label{sec:lexical_exp}

\begin{table}[h]
\centering
\caption{Comparison of lexical retrieval approaches with optional query augmentation on $1,000$ samples from the training set. Preprocessing includes lowercasing, punctuation removal, and subword tokenization.}
\label{tab:preprocessing-comparison}
\begin{tabular}{lccccc}
\toprule
 & \multicolumn{5}{c}{\textbf{Precision@k (\%)}} \\
 \cmidrule(lr){2-6}
\textbf{Approach} & \textbf{k=1} & \textbf{k=5} & \textbf{k=10} & \textbf{k=15} & \textbf{k=20} \\
\midrule
BM25 Baseline & 49.90 & 60.50 & 64.70 & 67.10 & 69.70 \\
BM25 + Preprocessing & 56.00 & 70.10 & 74.30 & 76.80 & 78.10 \\
Query Expansion + Preprocessing & 56.80 & \textbf{71.50} & \textbf{76.80} & \textbf{79.00} & \textbf{80.60} \\
Query Rewriting + Preprocessing & \textbf{57.80} & 70.90 & 75.10 & 77.50 & 79.60 \\
\bottomrule
\end{tabular}
\vspace{0.1cm}
\end{table}

To explore potential improvements in lexical retrieval, we experimented with two query reformulation strategies using the Gemma3 12B language model \cite{gemmateam2025gemma3technicalreport}. These methods aim to reduce the linguistic mismatch between informal social media posts and formal scientific abstracts.
\begin{enumerate}
\item \textbf{Query Rewriting}: Reformulating social media posts to correct grammar and match the formal language style of scientific abstracts while preserving the original query semantics (see Listing \ref{lst:rewrite_prompt}).
\item \textbf{Query Expansion}: Augmenting the original social media post with $2$-$3$ contextually relevant sentences to increase n-gram overlap with scientific abstracts (see Listing \ref{lst:expansion_prompt}).
\end{enumerate}
Among the evaluated methods, query expansion yielded the highest performance, achieving a Precision@20 of $80.6$\%, an improvement of $2.5$ percentage points over BM25 with preprocessing. Query rewriting also led to performance gains, with a Precision@20 of $79.6$\% (a $1.5$ percentage point improvement). 

However, both methods incur significant computational overhead due to the reliance on a large language model. Specifically, inference time increased by approximately a factor of 60. Furthermore, given that our pipeline includes a subsequent re-ranking stage, the marginal precision gains from these query reformulations diminish in the final results of the entire pipeline. This unfavorable cost-benefit trade-off renders these methods impractical for integration into the final pipeline, so we excluded them.

\begin{center}
\begin{minipage}{0.8\linewidth}
\begin{lstlisting}[style=promptstyle, caption={Prompt template to rewrite a social media post.}, label={lst:rewrite_prompt}]
Translate informal text into precise academic language, preserving
original meaning.

Transformation Guidelines:
- Correct the original tweet's spelling and grammar errors while maintaining its style
- Convert colloquial language to precise academic terminology
- Convert hastags into proper words
- Do not add anything new. Only correct the mistakes in the original tweet.

Output format:
Return a single string

Example:
Original Tweet: "Just saw amazin new study - mice w/ #Alzheimers showed
45% improvemnt in memory after new drug treatment!! Game changer for
#neurodegeneration research imo"

Output:
Just saw amazing new study - mice with Alzheimers showed 45% improvement
in memory after new drug treatment!! Game changer for
neurodegeneration research in my opinion

Transform the following tweet:
{tweet}
\end{lstlisting}
\end{minipage}
\end{center}

\begin{center}
\begin{minipage}{0.8\linewidth}
\begin{lstlisting}[style=promptstyle, caption={Prompt template to expand the social media post.}, label={lst:expansion_prompt}]
Translate informal text into precise academic language, according to the
transformation guidelines.

Transformation Guidelines:

First, correct the original tweet's spelling and grammar errors while
maintaining its style. Then transform the tweet into academic language
using these rules:

-Convert colloquial language to precise academic terminology
-Maintain semantic accuracy of the original message
-Use passive voice and objective scientific tone
-Eliminate informal expressions and subjective qualifiers
-Transform hashtags into their full, proper form (e.g., "#COVID19" -> "COVID-19 pandemic")
-Expand abbreviations and acronyms to their full forms
-Include key research terms that would appear in academic database searches
-Preserve all factual claims, statistics, and findings mentioned
-Structure as a concise academic abstract (2-3 sentences)

Output format:
Return a single continuous string with both versions separated by " || " as follows:
[Corrected Tweet] || [Academic Version]

Example:
Original Tweet: "Just saw amazin new study - mice w/ #Alzheimers showed
45% improvemnt in memory after new drug treatment!! Game changer for
#neurodegeneration research imo"

Output:
"Just saw amazing new study - mice with #Alzheimers showed 45%
improvement in memory after new drug treatment!! Game changer for
#neurodegeneration research in my opinion || A recent pharmacological
intervention demonstrated significant efficacy in an Alzheimer's
disease mouse model, with subjects exhibiting a 45% improvement in
memory function following administration of the novel compound.
These findings represent a potentially significant advancement in
neurodegenerative disease research, particularly regarding
therapeutic approaches for memory deficit amelioration in Alzheimer's
pathology."

Transform the following tweet:
{tweet}
\end{lstlisting}
\end{minipage}
\end{center}

\section{Embedding Model Fine-Tuning Details}\label{sec:details_embedding_ft}

To fine-tune the semantic embedding model, we initialize from \texttt{INF-Retriever-v1} \cite{infly-ai_inf-retriever-v1_nodate}, a transformer encoder pre-trained for dense retrieval tasks. Fine-tuning is performed by applying low-rank adaptation (LoRA) \cite{hu_lora_2022} to the query and value projection layers of the self-attention modules in the top $8$ transformer layers (layers $20$–$27$) using a rank $r=8$, scaling $\alpha=32$, and dropout of $0.1$.
The inputs are tokenized independently for queries (social media posts) and documents (title + abstract). The maximum sequence length is $8,192$ tokens, allowing for the processing of social media posts and documents without truncation.

We use the multiple negatives ranking (MNR) loss~\cite{henderson_efficient_2017}. Given a batch of $N$ query–document pairs, each query is trained to score highest on its corresponding document, while all other $N-1$ documents in the batch act as negatives.
We extract embeddings using \textit{last-token pooling}, which selects the hidden state of the final token in each sequence. Embeddings are $L_2$-normalized, and cosine similarity is computed via dot product.

We use AdamW \cite{loshchilov_decoupled_2019} as optimizer with a learning rate of $1 \times 10^{-5}$. We use $20$ linear warmup steps and then decay the learning rate to $0$ using a cosine scheduler. We train on $2$ A-100 GPUs using DDP with a per-device batch size of $4$ and $16$ gradient accumulation steps (resulting in an effective batch size of $64$).
We use gradient clipping with $\text{norm}=1.0$ and use FP16 mixed precision.
We evaluate \emph{retrieval quality} (i.e., run the vector store)  on the development set after each epoch by measuring Precision@100. The final model checkpoint is selected based on the best performance, which is obtained after epoch $2$.

\section{Comparison of Re-Ranking Models} \label{sec:rerankers}

\begin{table}[h]
\centering
\caption{Performance comparison of different re-ranking models using top $50$ semantic retrieval candidates on the development set. The chosen model and best scores are displayed in bold.}
\label{tab:reranker-comparison}
\begin{tabular}{lcc}
\toprule
\textbf{Re-ranking Model} & \textbf{MRR@1 (\%)} & \textbf{MRR@5 (\%)} \\
\midrule
\multicolumn{3}{c}{\textit{Baseline}} \\
\midrule
Semantic Retrieval & 61.50 & 67.56 \\
\midrule
\multicolumn{3}{c}{\textit{Traditional Cross-Encoders}} \\
\midrule
mxbai-rerank-large-v2 & 61.43 & 66.80 \\
bge-reranker-large & 62.50 & 67.54 \\
\midrule
\multicolumn{3}{c}{\textit{LLM-based Cross-Encoders}} \\
\midrule
\textbf{bge-reranker-v2-gemma} & 71.36 & \textbf{76.03} \\
bge-reranker-v2-minicpm [Layer 28] & 71.57 & 75.87 \\
bge-reranker-v2-minicpm [Layer 32]  & \textbf{71.79} & 76.02 \\
\bottomrule
\end{tabular}
\end{table}

We conducted a comparative evaluation of various re-ranking models on the development set to identify the most effective approach for our retrieval pipeline. The evaluated re-ranking models include traditional cross-encoders (\texttt{mxbai-rerank-large-v2} \cite{rerank2024mxbai}, bge-reranker-large \cite{bge_embedding}) and LLM-based re-rankers (\texttt{bge-reranker-v2-gemma} \cite{li_making_2023, chen_m3-embedding_2024}, \texttt{bge-reranker-v2-minicpm} \cite{li_making_2023, chen_m3-embedding_2024}), which use pre-trained language models as base for relevance scoring. The \texttt{bge-reranker-v2-minicpm} model supports layer-wise inference optimization, allowing computation to terminate at intermediate layers rather than processing through the full network.
We experimented with two different intermediate layer configurations, terminating after layer $28$ and layer $32$. We selected layer $32$ based on our preliminary experiment with $100$ samples across all available layers, which showed that layer $32$ achieved the best performance. Additionally, we included layer $28$, as this is recommended by the official BGE re-ranker repository. All re-rankers are evaluated on the development set using semantic retrieval candidates as input. As shown in Table \ref{tab:reranker-comparison}, LLM-based re-rankers outperformed traditional cross-encoders by a considerable margin. This performance gap likely stems from LLMs' extensive pre-training on diverse text corpora, enabling them to comprehend both formal and informal language patterns. Between the three LLM-based re-rankers, \texttt{bge-reranker-v2-gemma} achieved the best MRR@5 performance ($76.03$\% vs. $76.02$\% and $75.87$ \%). Although the margin is small, we selected \texttt{BAAI/bge-reranker-v2-gemma} as our final model.

\section{Data Augmentation for Semantic Retrieval}\label{sec:prompt_templates}
To enrich semantic retrieval, we experimented with two text augmentation strategies: hypothetical document embeddings (HyDE) \cite{gao_precise_2023} and additional documents (AD). Both methods leverage the Llama 3.2 7B model \cite{grattafiori_llama_2024} to generate auxiliary text representations.

For HyDE, we prompted the model to generate a hypothetical scientific article (title and abstract) based on a given social media post, aiming to bridge the domain gap between informal social media language and formal scientific discourse (see Listing~\ref{lst:hyde_prompt}). For AD, we augmented the document corpus by generating (1) a summary and (2) a synthetic social media post for each document. These variants were stored alongside the original document in the vector index (Listings~\ref{lst:ad_prompt_1} and~\ref{lst:ad_prompt_2}).

\begin{center}
\begin{minipage}{0.8\linewidth}
\begin{lstlisting}[style=promptstyle, caption={Prompt template to generate hypothetical document embeddings.}, label={lst:hyde_prompt}]
You are an expert in scientific research. Based on the following tweet,
generate a hypothetical scientific paper that includes only a title and
an abstract. The abstract should succinctly summarize the research
objective, methodology,  key findings, and conclusions.

Tweet: {tweet}

{format_instructions}
\end{lstlisting}

\begin{lstlisting}[style=promptstyle, caption={Summary Prompt Template for AD}, label={lst:ad_prompt_1}]
Summarize the following document:

Title: {title}
Abstract: {page_content}

Make sure to include keywords that are likely to be found later by a
search.

{format_instructions}
\end{lstlisting}

\begin{lstlisting}[style=promptstyle, caption={Tweet Prompt template to generate additional documents.}, label={lst:ad_prompt_2}]
Generate a hypothetical Twitter tweet about the following document:

Title: {title}
Abstract: {page_content}

Make sure it looks like a typical tweet from an average person and is
not too long.

{format_instructions}
\end{lstlisting}
\end{minipage}
\end{center}

\begin{table}
  \caption{Detailed performance comparison on the development set of semantic retrieval with additional text pre-processing.
  ``AD'' denotes additional documents, ``HyDE'' stands for hypothetical document embeddings. The best results per category are in bold.}
  \label{tab:performance_dev_detail}
  \begin{tabular}{lcc|cc}
    \toprule
     & \multicolumn{2}{c|}{\textbf{MRR@k (\%)}} & \multicolumn{2}{c}{\textbf{Precision@k (\%)}} \\
     \cmidrule(lr){2-3} \cmidrule(lr){4-5}
     \textbf{Approach} & \textbf{k=1} & \textbf{k=5} & \textbf{k=30} & \textbf{k=100} \\
    \midrule
    INF-Retriever-v1 & 58.66 & 65.21 & \textbf{86.50} & \textbf{92.04} \\
    \,\,+ HyDE & 48.89 & 56.06 & 80.37 & \textbf{87.44} \\
    \,\,+ AD & \textbf{60.73} & \textbf{66.69} & \textbf{87.09} & 90.87 \\
    \,\,+ HyDE + AD & 51.63 & 58.46 & 81.56 & 85.36\\
    \midrule
    INF-Retriever-v1 + Fine-tuning & 60.86 & \textbf{67.19} & \textbf{87.86} & \textbf{93.12} \\
    \,\,+ HyDE & 51.53 & 58.79 & 83.01 & 89.60 \\
    \,\,+ AD & \textbf{61.88} & \textbf{67.89} & \textbf{88.49} & 91.50 \\
    \,\,+ HyDE + AD & 53.44 & 60.56 & 83.60 & 87.32 \\
    \bottomrule
  \end{tabular}
\end{table}

The results on the development set are displayed Table~\ref{tab:performance_dev_detail}.
As discussed in Section~\ref{sec:results}, our primary objective for semantic retrieval is to ensure high precision, providing strong candidates for downstream re-ranking. We find that augmentation strategies offer modest improvements for off-the-shelf models but yield limited or no benefit when applied to the fine-tuned retriever. We hypothesize that the limited benefit observed from these augmentation methods stems from the fine-tuned model’s already high semantic fidelity, which reduces the marginal gains achievable through additional data augmentation. Therefore, these methods were excluded from the final pipeline.

\section{Hybrid Search using Elasticsearch} \label{sec:elastic_search}
In addition to our main retrieval pipeline, we explored a fully integrated alternative using Elasticsearch\footnote{\url{https://www.elastic.co}}. This system unifies indexing, retrieval, and ranking into a single framework, while still capturing both lexical and semantic signals.

We build an Elasticsearch pipeline closely mirroring our original architecture: it incorporates (1) a BM25 retriever with fuzzy matching, (2) a $k$-nearest neighbor (kNN) semantic retriever using our fine-tuned embedding model (configured with $k=50$ and $200$ candidates), and (3) a fusion stage based on reciprocal rank fusion (RRF) to combine results \cite{cormack_reciprocal_2009}. The RRF configuration uses a window size of $100$ and a rank constant of $20$, allowing it to integrate signals from both retrieval branches efficiently. Unlike our main system, which uses a cross-encoder for deep re-ranking, the Elasticsearch pipeline relies on this lightweight re-scoring mechanism.

\paragraph{Performance.}
\begin{table}[h]
\caption{Elasticsearch (ES) hybrid pipeline results on the development set. ``AD'' denotes additional documents, ``HyDE'' stands for hypothetical document embeddings. The best results are displayed in bold.}
\label{tab:Elasticsearch_results}
\centering
\begin{tabular}{lcc}
\toprule
\textbf{Configuration} & \textbf{MRR@1 (\%)} & \textbf{MRR@5 (\%)}\\
\midrule
ES & 60.53 & 66.69 \\
ES + HyDE & 45.99 & 53.56 \\
ES + AD & \textbf{63.72} & \textbf{69.35} \\
ES + AD + HyDE & 51.65 & 58.55 \\
\bottomrule
\end{tabular}
\end{table}

Similar to the evaluation of our custom pipeline described in Appendix~\ref{sec:prompt_templates}, we evaluate different variants of this Elasticsearch—based method, leveraging raw and extended documents, as well as with and without query expansion. Table~\ref{tab:Elasticsearch_results} presents the results obtained on the development set.

The best Elasticsearch configuration (ES + AD) achieves MRR@5 of $69.35$\%, slightly superior to our custom pipeline’s semantic retriever. However, it lags behind the full system with cross-encoder re-ranking (MRR@5 = $76.46$\%). This highlights the benefit of contextual re-scoring for fine-grained relevance. Nonetheless, the Elasticsearch-based approach remains a viable, scalable option for latency-sensitive applications.

\end{document}